\documentclass{osa-article}
\usepackage{subcaption} %need this package to get subfigures working
\usepackage{color,soul}
\usepackage{hyperref}
\journal{boe}
\articletype{Research Article}
\usepackage{lineno}
\begin{document}

\title{Designing and simulating realistic spatial frequency domain imaging systems using open-source 3D rendering software}

\author{Jane Crowley,\authormark{1,*} and George S.D. Gordon,\authormark{1}}

\address{\authormark{1}Optics \& Photonics Group, Department of Electrical and Electronic Engineering, University of Nottingham, Nottingham, United Kingdom}

\email{\authormark{*}jane.crowley@nottingham.ac.uk} %% email address is required

% \homepage{http:...} %% author's URL, if desired

%%%%%%%%%%%%%%%%%%% abstract %%%%%%%%%%%%%%%%
%% [use \begin{abstract*}...\end{abstract*} if exempt from copyright] 

\begin{abstract}
Spatial frequency domain imaging (SFDI) is a low-cost imaging technique that can deliver real-time maps of absorption and reduced scattering coefficients, offering improved contrast over conventional reflectance methods for imaging important tissue structures such as tumours. However, there are a wide range of imaging geometries that practical SFDI systems must cope with including imaging flat samples \emph{ex vivo}, imaging inside tubular lumen \emph{in vivo} such as in an endoscopy, and measuring tumours or polyps of varying shapes, sizes and optical properties. There is a need for a design and simulation tool to accelerate design and fabrication of new SFDI systems, and to validate performance under a wide range of realistic imaging scenarios. We present such a system implemented using open-source 3D design and ray-tracing software \emph{Blender} that is capable of simulating turbid media with realistic optical properties (mimicking healthy and cancerous tissue), a wide variety of shapes and size, and in both planar and tubular imaging geometries. Because Blender is a full end-to-end rendering package, effects such as varying lighting, refractive index changes, non-normal incidence, specular reflections and shadows are naturally accounted for, enabling realistic simulation of new designs. We first demonstrate quantitative agreement between Monte-Carlo simulated scattering and absorption coefficients and those measured from our Blender system.  Next, we show the ability of the system to simulate absorption, scattering and shape for flat samples with small simulated tumours and show that the improved contrast associated with SFDI is reproduced. Finally, to demonstrate the versatility of the system as a design tool we show that it can be used to generate a custom look-up-table for mapping from modulation amplitude values to absorption and scattering values in a tubular geometry, simulating a lumen. As a demonstrative example we show that longitudinal sectioning of the tube, with separate look-up tables for each section, significantly improves accuracy of SFDI, representing an important design insight for future systems.  We therefore anticipate our simulation system will significantly aid in the design and development of novel SFDI systems, especially as such systems are miniaturised for deployment in endoscopic and laparoscopic systems.
%The abstract should be limited to approximately 100 words. If the work of another author is cited in the abstract, that citation should be written out without a number, (e.g., journal, volume, first page, and year in square brackets [Opt. Express {\bfseries 22}, 1234 (2014)]), and a separate citation should be included in the body of the text. The first reference cited in the main text must be [1]. Do not include numbers, bullets, or lists inside the abstract.
\end{abstract}

%%%%%%%%%%%%%%%%%%%%%%%%%%  body  %%%%%%%%%%%%%%%%%%%%%%%%%%
\section{Introduction}
\label{sect:Intro}
\begin{figure}[!htbp]
     \centering
     \includegraphics[width=1\linewidth]{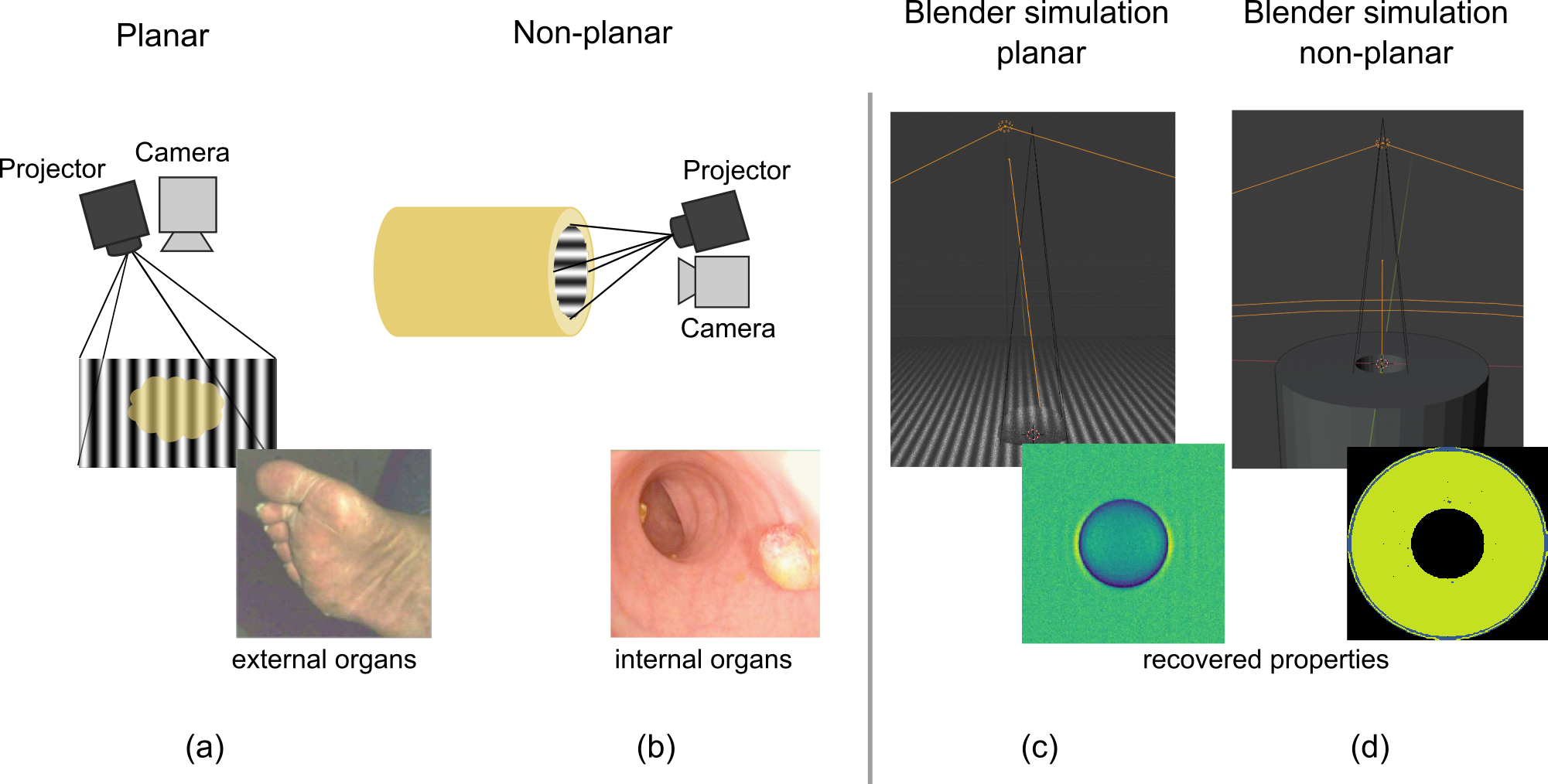}
     \caption{Future SFDI systems, especially those for \emph{in vivo} clinical use, may require significantly different geometries from conventional SFDI: a) conventional `planar' SFDI imaging geometry with projector at a small angle to flat sample, with real-world application of measuring diabetic foot shown in inset \cite{LeeDiabeticFoot2020}, b) SFDI operating in a tubular (lumen) geometry, that may be required for use in future endoscope systems where projection is no longer approximately flat, with example usage for imaging polyps in the colon shown inset \cite{Ribeiro2021}, c) screenshot of our \emph{Blender} SFDI model applied to a planar geometry, with reconstructed scattering properties of tumour like sample shown inset, d) a screenshot of our \emph{Blender} model applied to a non-planar tubular geometry, with reconstructed scattering properties shown inset.}
    \label{fig:ConceptFig}
\end{figure}

Optical properties, specifically scattering and absorption, and shape are important potential indicators of cancer within the gastrointestinal (GI) tract\cite{Holmer2007, Rex2019}. Conventional white light endoscopes and capsule endoscopes are the standard method of imaging the GI tract but  provide limited information about tissue properties that are hallmarks of a range of potential tumours\cite{Cummins2019}, leading to low five-year survival rates of oesphageal cancer (15\%\cite{CRUK}) and colon cancer (63\%\cite{ColonCancerStat}). SFDI is a well-established, low-cost imaging technique\cite{Dognitz1998}, with applications for imaging blood oxygenation\cite{FirstInHumanGioux}, burn depth\cite{Mazhar2014}, dental caries\cite{Bounds2021}, bowel ischaemia\cite{RodriguezLuna2022}, and indicators of cancer\cite{Angelo2017}. A range of commercial\cite{SFDImodulim} and research\cite{Gioux2019, Erfanzadeh2018, Applegate2020} SFDI systems are now available. However, these existing systems are almost exclusively designed for \emph{planar} imaging geometries, where the sample is flat and the camera and projector are located above it at near-normal incidence. However, many important clinical applications exhibit \emph{non-planar} geometries: for example imaging inside tubular lumen such as the GI tract, blood vessels, biliary system (Fig \ref{fig:ConceptFig}). SFDI imaging \emph{in vivo} in such organs is challenging due to miniaturisation needs, and because the surfaces are cylindrical, creating non-planar illumination conditions and sample geometries. This means that illumination and imaging may no longer be normal (or nearly normal) to the surface being imaged so different scattering behaviour will be observed \cite{Drezek1999}, and specular reflections will be altered.
To aid in the design of novel SFDI systems under these constraints, we have created an SFDI design and simulation tool in the open source $3$D modelling software \emph{Blender} (v 2.93) using the built-in ray-tracing engine Cycles (Fig \ref{fig:ConceptFig}). Cycles is a physically based path tracer, in which rays of light are traced from each camera pixel into the scene and either absorb into the world background, reflect, refract or reach their maximum bounce limit, typically $1024$ bounces. To increase accuracy, Cycles produces randomised light rays and averages the results from a singular pixel over time, analogous to a Monte Carlo simulation\cite{IntrotoMC}. Cycles simulates volume scattering inside objects using a Henyey-Greenstein Phase function, which is commonly also used in Monte-Carlo simulations of tissue \cite{Wang1995,BlenderSource}. \emph{Blender} has previously been used for three-dimensional shape measurement of additive manufacturing parts with complex geometries\cite{Eastwood2020a}, for the development anatomically accurate meshes to use in Monte Carlo light simulations\cite{Zhang2022}, and for the generation of SFDI image data sets to train neural networks \cite{JCrowley_SPIE, Osman2022}.  By using Blender for both geometry specification (i.e. design) and simulation (via ray-tracing with Cycles), we are able to simulate realistic optical properties and geometries while naturally accounting for realistic features of SFDI systems such as stray light, specular reflections and shadows.

\begin{figure}[!htbp]
    \centering
    \includegraphics[width=1\linewidth]{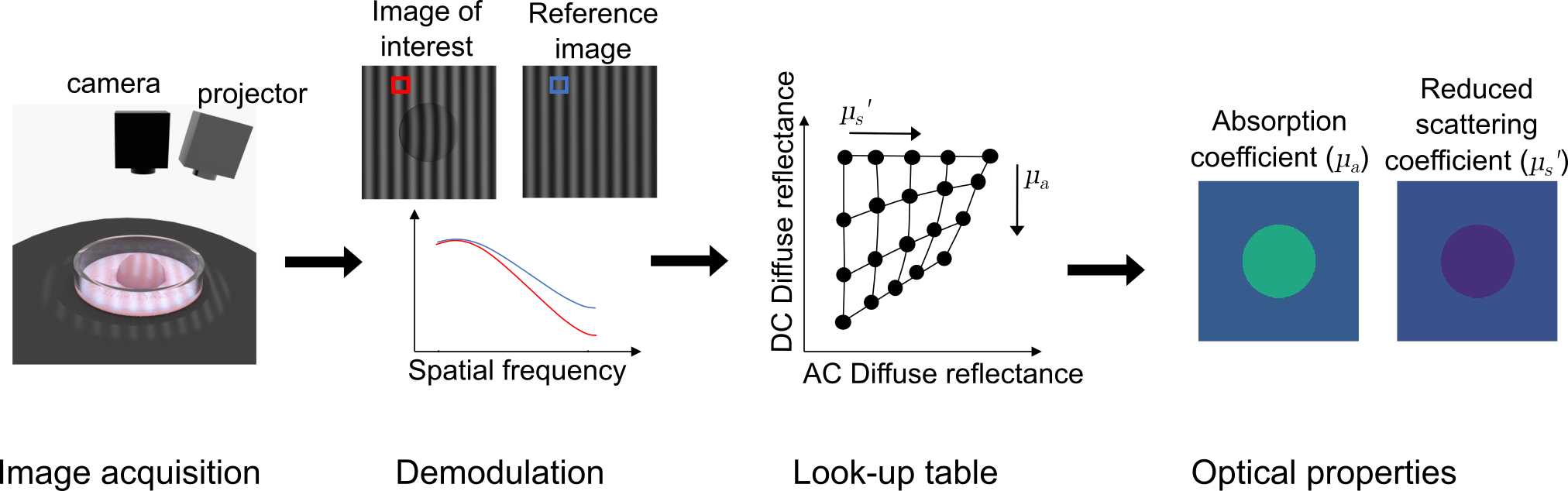}
    \caption{Conventional SFDI process: A sample is illuminated with a sinusoidal pattern then an image is captured. This is then demodulated and compared with a reference material of known optical properties. The optical properties of the material of interest are then estimated using a look-up table.  Finally, maps of absorption and reduced scattering coefficient are produced.}
    \label{fig:SFDI}
\end{figure}

Conventional imaging in the spatial frequency domain is demonstrated in Fig \ref{fig:SFDI} and consists of projecting a known set of structured illumination patterns onto a sample at a small angle ($\lesssim 6^{\circ}$) to the normal to minimise specular reflections recorded by the camera \cite{Cuccia2005}. The structured illumination set typically consists of 2D sinusoids at 3 equispaced phase offsets but recent work has shown the successful use of randomised speckle patterns as an alternative illumination scheme \cite{Chen2021}. The amplitude of the projection pattern is modified by the absorbing and scattering nature of the sample. A camera placed orthogonal to the sample captures images which are processed to determine the modulation amplitude of the reflected illumination pattern as a function of spatial frequency, via the equations\cite{Cuccia2009}:
\begin{equation}
        M_{AC}(x_i, f_x) = \frac{\sqrt{2}}{3}\sqrt{(I_1(x_i) - I_2(x_i))^2 + (I_2(x_i) - I_3(x_i))^2 + (I_3(x_i) - I_1(x_i))^2}
    \label{MAC}
\end{equation}
\begin{equation}
    M_{DC}(x_i) = \frac{1}{3}\left(I_1(x_i) + I_2(x_i) + I_3(x_i) \right)
\end{equation}    
where $I(x_i)$ is intensity of each pixel (position $x_i$) value in the captured images for sinusoidally modulated illuminations with a spatial phase shift of $0^{\circ}$ ($I_1$), $120^{\circ}$ ($I_2$) and $240^{\circ}$ ($I_3$). These results can also be obtained from a single image, termed \emph{single snapshot of optical properties (SSOP)}, by the use of Fourier-domain filtering \cite{Vervandier2013} or convolutional neural networks \cite{Chen2019} to separate the AC and DC components. In either approach, the AC and DC modulation amplitudes need to be calibrated with the modulation transfer function (MTF) of the system in order to produce \emph{diffuse reflectance} values that represent the next intermediate step toward obtaining scattering and absorption.
Conventional calibration for the MTF is achieved by imaging a reference material of known optical properties and computing diffuse reflectance values for these properties using a light propagation model: Monte Carlo simulation or the Diffusion Approximation\cite{Cuccia2009, Flock1989}. The difference between the computed and measured diffuse reflectance is used to infer the MTF, which can then be applied to obtain diffuse reflectance values from the modulation amplitudes of the sample of interest.  Finally, the absorption, $\mu_a$, and reduced scattering, $\mu_s\prime$, coefficients are then determined via a look-up table (LUT) generated from the chosen light-propagation model. An alternative method of generating this LUT is to measure a large data set of materials of known optical property values and interpolate between values to create an LUT unique to the imaging system in use. This has been achieved previously by calculating the reflectance and modulation of each material by comparison to a reflectance standard \cite{Erickson2010}.

Another important property that SFDI imaging can extract is height information, i.e. shape in flat geometries. This can be done via fringe projection profilometry\cite{Takeda1983, Su2001}. A structured illumination pattern is projected onto a sample of interest and a camera captures how the pattern is distorted by the presence of the sample to determine sample shape \cite{Angelo2017}. This information can be useful in clinical settings for quantifying the size and shape of polyps, which is linked to their pathology \cite{Rex2019}. Previous studies have shown that structured illumination  significantly reduces error in shape measurement compared to visual assessment, and is comparable to measurement with biopsy forceps and ruled snare \cite{Visentini-Scarzanella2018}. Combining shape information with optical property information would give valuable diagnostic information to the clinician to determine optimal treatment for patients.

Developing an SFDI system to determine optical properties and shape in clinical environments, both \emph{ex vivo} and \emph{in vivo}, has many challenges associated with it, such as determination of optimum illumination source placement and determination of optimum illumination patterns. Here, we present a design and simulation system using free, open-source 3D modelling and rendering package \emph{Blender}, that can implement SFDI for simulation of absorption, scattering and shape. We first show how to use \emph{Blender} to model a customisable scattering and absorbing material with in-built \emph{Blender} material nodes. We then show how to construct a virtual characterisation system for the absorption density, $A_{\rho}$, and scattering density, $S_{\rho}$, of this material using two approaches: a double integrating sphere (DIS)\cite{Pickering1993} and an SFDI system. For both approaches, we validate the accuracy of retrieved optical properties and show how this can improved by generating an empirically derived LUT from the DIS in-situ data. Next, we present two illustrative example use cases for our system.  First, we show that the simulated SFDI system enables reconstruction of scattering, absorption and shape of planar geometry samples mimicking cancerous and pre-cancerous conditions such as squamous cell carcinoma and Barrett's Oesophagus respectively. Second, we demonstrate, for the first time, a novel illumination scheme tailored for non-planar, tubular geometries (such as inside a lumen) where the spatial frequency is constant throughout the length of the tube such that the optical properties can be accurately obtained. To improve accuracy, we longitudinally section the tube and create separate look-up tables for each section, a straight-forward task in our system. We show that this customised illumination can detect changes in absorption and scattering properties within a tube of biologically relevant material, providing a potential design for future SFDI systems.

\section{Methods}
\subsection{Material simulation}
\label{subsec:MaterialSimulation}
The use of 3D modelling or CAD software to simulate conventional optical imaging with a light source, 3D objects and a camera is well-established. The key challenges for designing an SFDI simulation system in such software is ensuring accurate, calibrated simulation of optical scattering and absorption, and designing appropriately structured illumination patterns.

To achieve the most physically realistic ray-traced renderings in \emph{Blender}, some optimisation of the render settings is required. Within the ray-tracing engine Cycles, which physically traces rays of light, the maximum number of bounces a light ray can travel before the simulation terminates can be set. We set this value to $1024$, the highest allowed. The number of samples to render per pixel in the image was set to $1000$. Clamping of direct and indirect light, which limits the maximum intensity a pixel can have, was disabled by setting both to $0$. Colour management, which is typically used to make visually appealing images but introduces unwanted artefacts such as gamma correction, was disabled by setting the display device to \emph{`None'}. View transform was set to \emph{`Standard'} to ensure no extra conversions were applied to the resulting images. The sequencer, which sets the the colour space, was set to \emph{`Raw'} to avoid unwanted colour balancing or further gamma correction. For all images rendered, the camera exposure is adjusted to avoid saturation while maximising power of detected signal, but the images must then have their intensities corrected by following the equation:
\begin{equation}
    I_{output}(x, y) = I_{render}(x, y)\times 2^{t_{exposure}}
    \label{eqn:ExposureEqn}
\end{equation}
where $I_{output}$ is the exposure-corrected intensity we required, $I_{render}$ is the raw value obtained following the render, and $t_{exposure}$ is the exposure setting.

Previous work has used a weighted mixture between transparent, sub-surface scattering and absorbing materials to create a composite material with the desired optical properties \cite{JCrowley_SPIE}. Though this approach works in many realistic operating regimes, it is limited because the sub-surface approximation applies only at surfaces and not in the entire material volume. Here, we therefore model the material more accurately using a volume shader, instead of surface shader, exploiting Blender's built-in volume absorption and volume scattering shaders. The absorption and scattering was varied by changing the density parameters of the nodes, $A_{\rho}$ and $S_{\rho}$ respectively. The anisotropy, \emph{g}, in the volume scatter node was set to $0.8$, in line with the anisotropy values measured of the tissue within the GI junction\cite{Holmer2007} and with the value of \emph{g} set in the Virtual Photonics Monte Carlo simulation software\cite{MCCL2022}, which was used to generate a LUT. When capturing image data, we extract the red channel of the RGB colour images.  However, because Blender supports tri-colour operation it can also provide physically realistic scattering at green and blue wavelengths if desired.

In order to use a LUT generated from a Monte-Carlo simulation or the Diffusion Approximation, the semi-infinite thickness requirement must be met \cite{Contini1997}. To set an appropriate thickness for the material to meet this property, a red sphere was placed behind the material and the parameters were varied until the sphere was not visible, i.e. the material was not transparent for a fixed thickness. For a material of $2$ m thickness, this red sphere was not seen for $A_{\rho} \geq 2$ when $S_{\rho} = 0$, and for $S_{\rho} \geq 3$ when $A_{\rho} = 0$.   These are therefore the limitations of the material in our system when used for SFDI. However, it is noted that this limitation could be circumvented by using an empirically derived LUT that is calibrated to a particular physical thickness. 

Our aim is to create a simulation of an SFDI system with biologically relevant samples, and so we have identified two disease states relevant for detection of cancer in the upper GI tract that have distinctive scattering and absorption properties: squamous cell carcinoma (SCC) and Barrett's Oesophagus (BO). SCC is formed of neoplastic cells invading the submucosal layer of tissue\cite{Sweer2019}. For simplicity, we model SCC with a spheroid but the flexibility of Blender allows for generation of arbitrary shapes. We modelled tumour spheroids using sphere meshes scaled to be $40$ mm diameter. We note that the \emph{`scale'} parameter of the object in Blender should be reset when the desired size is reached to ensure proper behaviour with regard to scattering length scales along different dimensions. 

BO can be a pre-cursor to oesophageal adenocarcinoma, and occurs when the epithelium of the oesophagus begins transforming into a structure mimicking the the lining of the stomach\cite{Anaparthy2014}. To examine contrast, two materials were placed adjacent to one another: one with the optical properties of healthy oesophageal tissue and the other with the optical properties of BO. To simulate realistic gastrointestinal imaging, we consider two imaging geometries. The first simulates an `up-close' view of a tumour on the wall of a large lumen and can be approximated by a flat geometry.  However, to identify such structures during a typical endoscopy or to examine such structures in a smaller lumen, it is also necessary to consider a tubular geometry with a wide field-of-view. We therefore also consider the scenario of an SFDI system pointing down a tube, shown in Figure \ref{fig:ConceptFig}b.

\subsection{Calibration of material optical properties}
\label{subsect:Calibration of material optical properties}
For SFDI measurements, a reference material of known optical properties is required to correctly calibrate the system response (as discussed in Sect \ref{sect:Intro}). This requires determining the relationship between the material parameters in \emph{Blender} and the recovered absorption and reduced scattering coefficients. This can be done directly with an SFDI system through a `trial and error' approach -- the material parameters are adjusted until plots of diffuse reflectance vs. spatial frequency agree with theoretically derived curves \cite{JCrowley_SPIE}. However, this approach is imprecise and laborious. We therefore developed a more accurate approach which involves simulating a double integrating sphere (DIS) system in \emph{Blender} \cite{Pickering1993}. This system consists of two hollow spheres, termed the `reflectance' sphere and  `transmittance' sphere, each with $100$ mm diameter and $10$ mm wall thickness. The material of these spheres is set to be highly reflective using the diffuse bi-directional scattering distribution shader with $0$ roughness and colour of pure white. The reflectance sphere has an entry port and an exit port, with the sample located at the exit port, which are rectangular in shape with a $10$ mm side length. The transmittance sphere has only an entry port, where the sample is located, of the same shape and size as the reflectance sphere exit port. The sample placed at the exit (sample) port of the reflectance sphere and the entry (sample) port of the transmittance sphere has a thickness of $1$ mm. The material of the sample is that of the material described in Sect. \ref{subsec:MaterialSimulation}. 

The input light source is a spot light of power $5$ W, with a beam radius of $0.5$ mm and a spot size of $6^{\circ}$. The light is placed at the entry port of the reflectance sphere. Cameras were placed at the base of each of the spheres to act as detectors, with all pixels summed together (i.e. integrated over the detector area) to give a power value. For our initial tests, only the red channel is considered. A baffle is placed between the sample ports and the cameras to block specularly reflected light from the sample entering the camera detector. To perform normalisation, it is necessary to collect transmission and reflectance measures with a reflectance standard sample, no sample and with the light beam blocked.  The reflectance standard sample is simulated using the diffuse bi-directional scattering distribution function (BSDF) shader in \emph{Blender} with a purely white colour and roughness set to $0$. It is assumed that the normalised reflectance of this material is $0.9$. For each captured image, the camera exposure was varied until the average intensity was approximately in the middle of the 0-255 range (i.e. 8-bit colour).  This exposure was noted and corrected for using Eqn \ref{eqn:ExposureEqn}.

To determine the reduced scattering and absorption coefficients, a series of images is taken in the reflectance sphere and the transmittance sphere, and the normalised reflectance and transmittance are calculated respectively for varying sample material properties using the equations:
\begin{equation}
    M_R = r_{std} \frac{R_2(r_s^{direct}, r_s, t_s^{direct}, t_s) - R_2(0,0,0,0)}{R_2(r_{std}, r_{std}, 0,0) - R_2(0,0,0,0)}
\end{equation}
\begin{equation}
    M_T = \frac{T_2(r_s^{direct}, r_s, t_s^{direct}, t_s) - T_2(0,0,0,0)}{T_2(0,0,1,1) - T_2(0,0,0,0)}
\end{equation}
where $r_{std}$ is the normalised reflectance of the reflectance standard, $R_2(r_s^{direct}, r_s, t_s^{direct}, t_s)$ and $T_2(r_s^{direct}, r_s, t_s^{direct}, t_s)$ are reflectance and transmittance measurements respectively when the sample material is in place, $R_2(r_{std}, r_{std}, 0,0)$ is a reflectance measurement when the standard reflectance sample previously described is in place of the material, $R_2(0,0,0,0)$ is a reflectance measurement when there is no sample present and the transmittance sphere is removed, $T_2(0,0,1,1)$ is a transmittance measurement when light passes straight through the reflectance sphere when no sample is present into the transmittance sphere and $T_2(0,0,0,0)$ is a transmittance measurement when the incident beam is blocked and there is no sample in the port. These normalised values were then input into an inverse adding doubling (IAD) algorithm to determine the optical properties\cite{Prahl1993}. 

We used nine data points in the ranges $A_{\rho}: 1-100$ and $S_{\rho}: 5000-20000$. These values were chosen as when input into the IAD algorithm, they gave optical properties within our range of interest. We captured images in our SFDI set-up for these same material parameters, with a camera placed $0.5$ m above the sample of interest and a $5$ W spot light source, acting as the projector, placed at a $4^{\circ}$ offset to the camera to reduce any specular reflections. The camera and projector were placed at the same height from the sample, at $0.035$ m apart. The optical properties in the up-close flat geometry were calculated using two different LUTs: a Monte Carlo generated LUT and an empirically-derived LUT. 

\subsubsection{Monte Carlo LUT}
\label{subsubsec:Monte Carlo LUT}
The Monte Carlo (MC) LUT was generated using Virtual Photonics Monte Carlo simulation software \cite{MCCL2022}, with absorption coefficients ranging from $0-0.5$ mm$^{-1}$, and reduced scattering coefficients ranging from $0-5$ mm$^{-1}$. The optical properties of the nine material values were calculated using a reference material of $A_{\rho} = 1$ and $S_{\rho} = 20000$ with the corresponding reference optical properties determined from the IAD algorithm for this specific material.  
\subsubsection{Empirically-derived LUT}
\label{subsubsec:Empirically-derived LUT}
The empirically-derived LUT is able to correct for discrepancies between the SFDI and IAD measurements, which arise from the way anisotropy is implemented when mixing the absorption and scattering materials, leading to slight deviations in effective anisotropy constant ($g$). For our first simulation of an `up-close' view of the wall of a large lumen, approximated as a flat geometry, we generated a modulation vs reflectance LUT as described by \emph{Erickson et. al.} \cite{Erickson2010}. We started with the same nine data points as before, captured the modulation and reflectance of these densities, and then did a first linear interpolation between these data points to increase the LUT from 9 data points to 100 $\times$ 100 data points, improving granularity of final optical properties. A second interpolation step, this time using bicubic interpolation, was then carried out to determine the optical properties of a sample of interest. We also performed an inverse calculation, such that if one has desired optical properties of interest, they may determine what $A_{\rho}$ and $S_{\rho}$ are required to produce these optical properties. For our second simulated situation of a wide field-of-view tubular geometry, we generated a LUT of AC modulation amplitude (normalised to reference) vs DC modulation amplitude (normalised to reference). This type of LUT that uses normalisation was chosen because in the tubular geometry there is a difference in intensity from the distal to proximal end of the tube, and therefore large intensity variation within the material. Dividing by the reference material with similar variation in intensity reduces this intensity variation in the resultant image and makes for a more accurate optical property calculation. 

\subsection{Robust shape determination}
In addition to measuring optical properties, we would also like to exploit the fringe-projection approach to reconstruct 3D shape via fringe profilometry. To do this, we will consider a fringe projection pattern of the form:
\begin{equation}
    \psi(x,y) = \sin(\omega y + \phi) 
\end{equation}
where $\omega = 2\pi f$ is the angular frequency of the projected pattern with spatial frequency $f$. The sinusoidal pattern must be rotated $90^{\circ}$ from the optical property measurements (which are of the form $\psi(x,y) = \sin(\omega x + \phi)$) such that the fringes show maximum sensitivity to surface variations \cite{Gioux2009}. This is because a change in vertical height now corresponds to a displacement in one axis from the centre of the projector, which in turn results in a phase shift of the sinusoid.  This is a consequence of the small angle of the projector relative to the camera. If this displacement were instead along the fringes there would be no shift in phase observed.

Typically, a single fringe image may be used to reconstruct height but, as with SSOP, this may require spatial filtering and hence incur a reduction in resolution. While single-shot methods may offer a speed advantage, their noise performance is typically worse. We therefore use a generalised approach for using $N$ phase-shifted images to reconstruct height maps \cite{PhaseShiftZuo2018}. If the geometry of the system is precisely known, this phase can then be converted to height for each pixel in the image via the equation\cite{Takeda1983}:
\begin{equation}
    h(x, y) = \frac{l_0 \Delta\phi(x, y)}{\Delta\phi(x, y) - 2\pi f_0 d}
\end{equation}
where $l_0$ is the distance from the projector to the reference material, $\Delta\phi$ is the phase difference between the actual phase (calculated) and the phase of the background reference plane, $f_0$ is the spatial frequency of the projected pattern and $d$ is the separation distance of the projector and camera. Because of the geometrical assumptions made in mapping phase to height, this approach cannot be straightforwardly applied to non-planar geometries for shape reconstruction. In non-planar geometries, reconstruction of exact physical height could instead be approximately deduced by comparison with a reference phantom, e.g. perfectly straight tube for a lumen geometry, or by applying advanced techniques such as deep-learning \cite{Feng2019}.

\subsection{Development of projection pattern for tubular geometry}
\label{subsec:ProjPatternTube}
\begin{figure}[!htbp]
    \centering
    \includegraphics[width=1\linewidth]{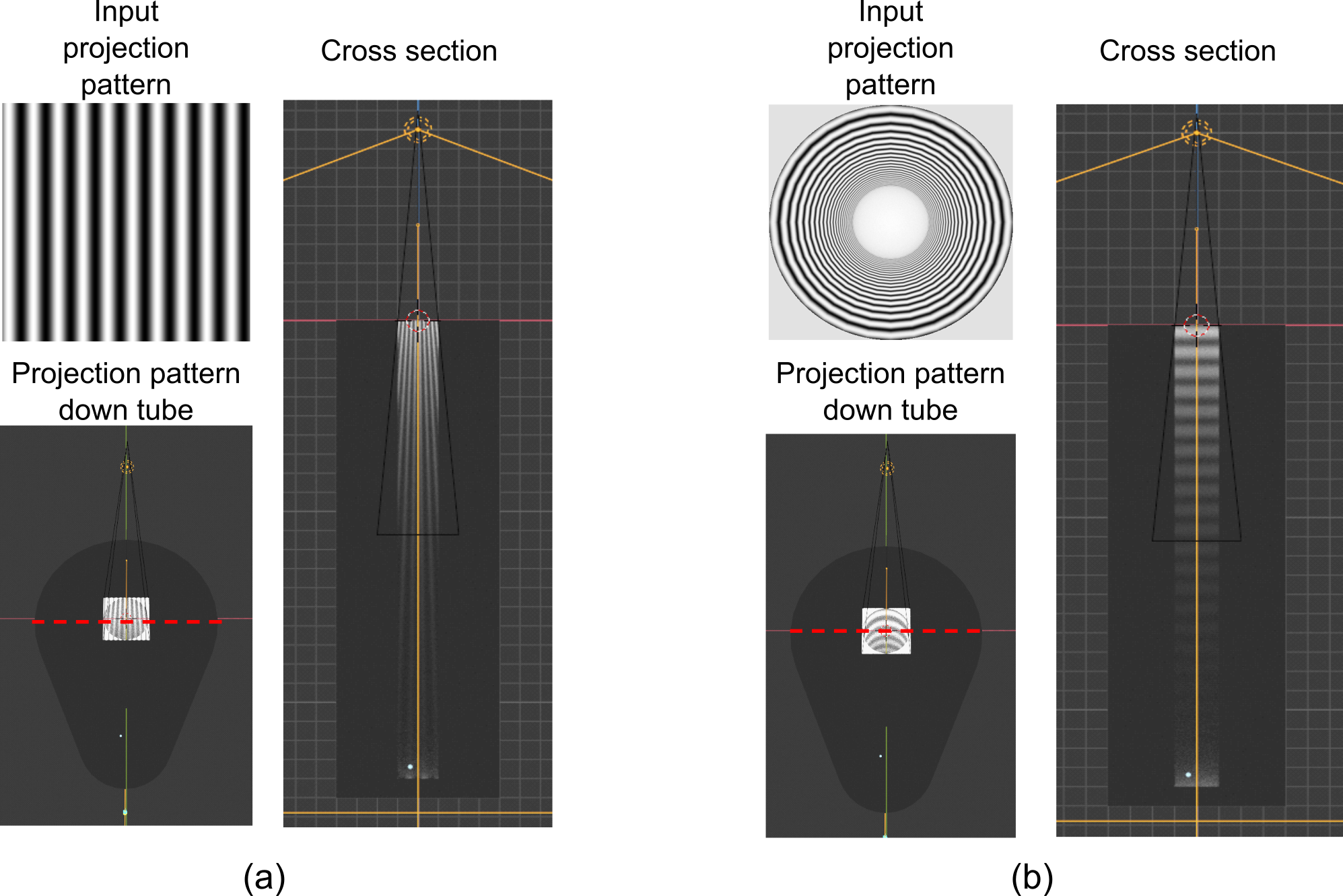}
    \caption{Comparing tubes with (a) planar sinusoidal pattern and (b) our novel illumination pattern. Top left shows image being projected from projector (highlighted yellow in complementary images). Bottom left shows view of shown projection pattern down tube with red dashed line indicating where tube is cut to view cross section on right panel}
    \label{fig:Tube_multiple}
\end{figure}

Conventional SFDI systems with planar sinusoidal projections are suitable for planar samples, such as \emph{ex vivo} resected tissue specimens. However for \emph{in vivo} use, an SFDI system would typically need to be operated inside a tubular lumen, e.g. the gastrointestinal tract if in an endoscope. Using our \emph{Blender} simulation it is very simple to explore such a situation. We began by simulating a tube of length $250$ mm with an outer diameter of $80$ mm and an inner diameter of $20$ mm. A $120$ mW spot light source was placed at a distance of $100$ mm from the top of the tube and projected a flat sinusoidal pattern down the tube. This naive approach creates a non-uniform spatial frequency pattern throughout the length of the tube which makes reconstructing accurate optical properties challenging (see Fig \ref{fig:Tube_multiple}a). Therefore, we developed a process to create a more suitable illumination pattern for other imaging geometries and demonstrated for the test case of a tube.
First, the material of the tube was set to be highly reflective using a pre-existing material node of diffuse BSDF with a roughness of $0$ and a shade of pure white. Next, the surface of the tube was `unwrapped' within \emph{Blender} using the UV mapping tool, resulting in a flattened map of the inside of the tube. A sinusoidal pattern of the desired phase and spatial frequency was then applied to this flat surface. Once applied, the material is then wrapped, such that the inside of the tube now has a uniform spatial frequency throughout its length. $1$ W light sources were placed equally throughout the tube such that the illumination intensity is uniform looking down the tube at the top. Here, we evenly distributed 40 point sources down the $250$ mm tube. A camera placed $110$ mm above the top of the tube then captured an image of the concentric circle illumination pattern. This image was then exported to \emph{Python} where a normalisation was applied to ensure that the sinusoid pixel values vary across the maximum range for projection ($0-255$). This process was carried out for sinusoidal patterns of a fixed spatial frequency at 3 different phase shifts. 

Then, the material of the tube can be reverted to the original scattering/absorbing material described in Sect \ref{subsec:MaterialSimulation}. Finally, the normalised images of the patterned tube are then used as the new projection patterns, which are projected onto the tube with a $5$ W light source.  The new projection pattern and resultant tube cross section is shown in Fig \ref{fig:Tube_multiple}b. This process can be considered a `pre-distortion' of the projected pattern to produce more uniform spatial frequencies and could alternatively be computed using analytically-derived formulae, or by direct inverse computation using a ray-tracing engine.  However, our approach here is very simple to implement and is highly effective. These modified projection patterns can then be used for SFDI imaging as there is a now a uniform spatial frequency pattern within the geometry length.

However, the tubular geometry inherently allows less light to reach the distal end of the tube and also less light to be reflected back as only a small range of angles can escape the tube via the opening. The projector placement, at a large angle to the normal of the tube surface, also creates different incidence angles along the length of the tube. The empirically-derived LUT approach still therefore produced substantially different optical properties along the length of the tube. As the LUT is constructed by taking the mean of pixels on the tube wall, this is an inevitable result using a single LUT for the whole tube. To improve these results, we developed a sectioning approach. Instead of using a single LUT as before, we sectioned the tube (length wise) into five different longitudinal subsections and used a different LUT for each section. The five sections were selected as regions that showed a mean intensity difference $>10$ relative to other sections.

\section{Results and Discussion}
%summarise what was done
%list novelties
%discuss limitations within each subsection
\subsection{Material simulation}
In the DIS simulation, $A_{\rho}$ and $S_{\rho}$ were varied, the reflectance and transmission values obtained, and input into the IAD algorithm which computes the optical properties of the material. This was repeated until optical properties within our range of interest were obtained. Following this process, it was found that realistic ranges for these parameters were $1\leq A_{\rho} \leq 100$ and $5000\leq S_{\rho} \leq 20000$. We selected material of $A_{\rho} = 1$ and $S_{\rho} = 20000$ with optical properties $\mu_a = 0.026 \textrm{ mm}^{-1}$ and $\mu_s \prime = 4.75 \textrm{ mm}^{-1}$ to be the reference material for the SFDI measurements using a Monte Carlo generated LUT as described in Sect \ref{subsubsec:Monte Carlo LUT}. 
\begin{figure}[!htbp]
    \centering
    \includegraphics[width=1\linewidth]{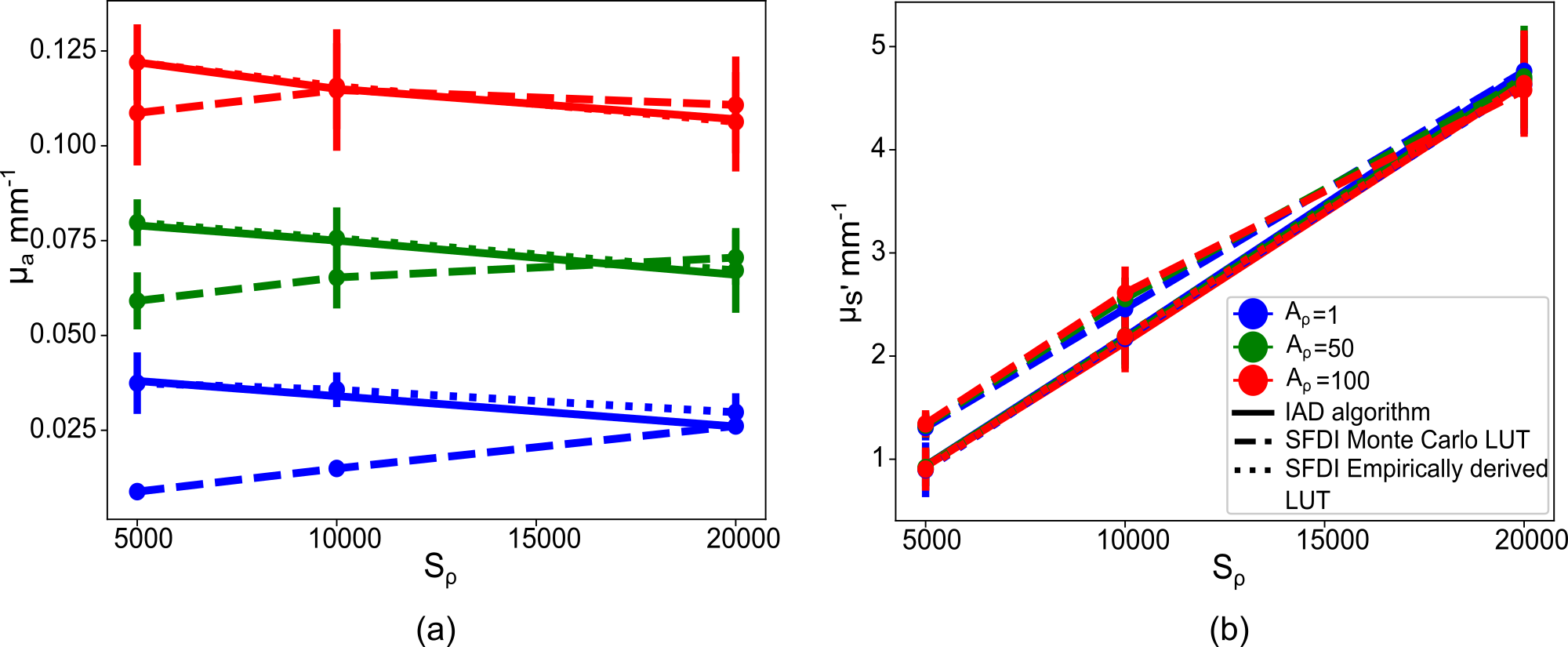}
    \caption{(a) Absorption and (b) reduced scattering coefficient vs scattering density, $S_{\rho}$, calculated for varying absorption densities, $A_{\rho}$, via IAD algorithm (solid line), SFDI Monte Carlo LUT (dashed line) and SFDI empirically derived LUT (dotted line). The error bars represent the standard deviation across the calculated $500\times500$ pixel optical property map.}
    \label{fig:SFDIplots}
\end{figure}
The results from the SFDI measurements are shown in Fig \ref{fig:SFDIplots}. We note from Fig \ref{fig:SFDIplots}(a) that there is a discrepancy between the results from the IAD and the SFDI Monte Carlo LUT calculations, particularly at low $S_{\rho}$ values. We also note a small cross-coupling between absorption and scattering: increased scattering density reduces absorption coefficient. The reason for the first of these two irregularities can be seen by examining the implementation of the material in Blender. The scattering and absorbing materials are combined using an `add' shader which equally weights both materials and adds them together. Because the absorbing component of the material does not display anisotropy, unlike the scattering component, the combination of the two components slightly modifies the `effective' anisotropy of the whole material to a value different from the nominal $g$ of the scattering component.  This is the cause of the small discrepancy between the two methods of measuring scattering. The cross-coupling between scattering and absorption may arise in part because increased absorption reduces the accuracy of scattering measurements as there will be fewer `scattering' events simulated for each ray before it is absorbed. The effect observed here is comparatively small and so for the purposes of designing SFDI systems may be neglected.  However, we note that such cross-coupling is observed, often more strongly, when imaging experimental phantoms.

To account for these discrepancies, we introduce the empirically-derived LUT described in Sect \ref{subsubsec:Empirically-derived LUT} with resultant calculated optical properties displayed in Fig \ref{fig:SFDIplots}. Using this we are able to select optical properties we wish to simulate and input the associated $A_{\rho}$ and $S_{\rho}$ values. We can then simulate materials of various shape and give them specific optical properties, which is ultimately the most relevant feature required for designing and testing new SFDI systems.

\subsection{Simulation of typical gastrointestinal conditions in up-close planar geometry}
\begin{figure}[!htbp]
    \centering
    \includegraphics[width=1\linewidth]{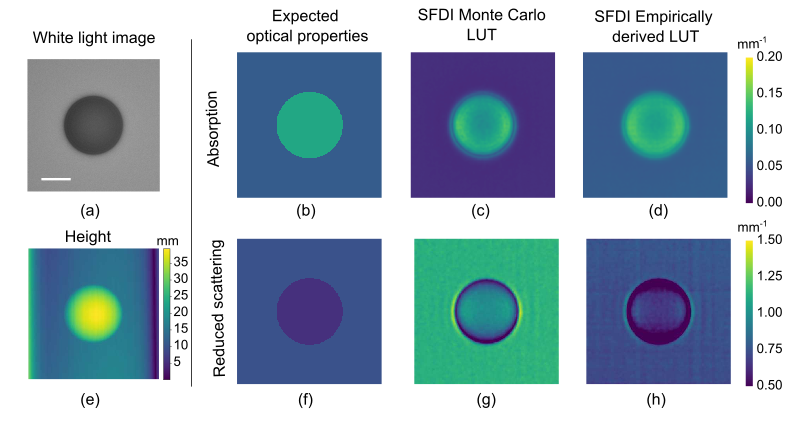}
    \caption{Simulated squamous cell carcinoma as a spheroid on a background of healthy oesophageal tissue showing (a) white light image and (e) reconstructed height map with (b) expected absorption coefficient, $\mu_a$, (c) $\mu_a$ recovered with MC LUT (d) $\mu_a$ recovered with empirically derived LUT (f) expected reduced scattering coefficient, $\mu_s\prime$, (g) $\mu_s\prime$ recovered with MC LUT and (h) $\mu_s\prime$ recovered with empirically derived LUT. Scale bar = 20mm.}
    \label{fig:SCC_planar}
\end{figure}

Fig \ref{fig:SCC_planar} shows the optical property and height maps generated for a $40$ mm diameter simulated polyp, with an absorption coefficient higher than that of surrounding healthy tissue and a reduced scattering coefficient lower than that of surrounding healthy tissue, simulating squamous cell carcinoma. At $635$ nm, the absorption coefficient of squamous cell carcinoma is $0.12$ mm$^{-1}$, which is much greater than that of healthy oesophageal tissue, $0.058$ mm$^{-1}$, and the reduced scattering coefficient of $0.64$ mm$^{-1}$, is less than that of healthy oesophageal tissue, which is typically $0.75$ mm$^{-1}$\cite{Sweer2019}. To recreate this behaviour we therefore simulated healthy background tissue with $A_{\rho} = 25$ and $S_{\rho} = 4236$, and we simulated a polyp with $A_{\rho} = 96$ and $S_{\rho} = 3721$. The simulated polyp has a height (from the base material) of $40$ mm. Fig \ref{fig:SCC_planar}e shows a successful height map generation from fringe profilometry measurements. Fig \ref{fig:SCC_planar}c,d,g,h demonstrate successful recovery of optical properties, demonstrating the success of this method to simulate and image objects. We note that the empirically derived LUT produces results closer to the expected values, which is because it accounts for some of the discrepancies in our tissue simulation as described earlier.  However,the Monte Carlo LUT still provides high contrast between the squamous cell carcinoma and background, which is arguably more important for wide-field diagnostic applications.  We note that because the surface profile information is available, the optical property accuracy may be improved by the addition of surface profile correction for optical property determination\cite{Gioux2009}. 

\begin{figure}[!htbp]
    \centering
    \includegraphics[width=1\linewidth]{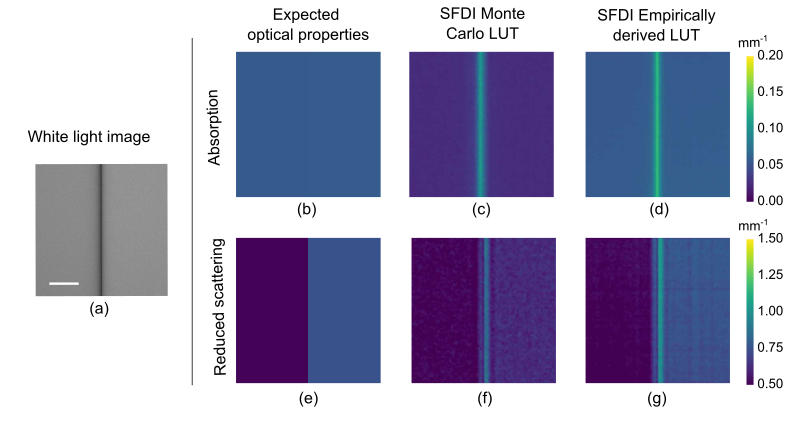}
    \caption{Simulated Barrett's Oesophagus with mild chronic inflammation (left half of sample) adjacent to healthy oesophageal tissue (right half of sample) showing (a) white light image (b) expected absorption coefficient, $\mu_a$, (c) $\mu_a$ recovered with MC LUT and (d) $\mu_a$ recovered with empirically derivied LUT (e) expected reduced scattering coefficient, $\mu_s\prime$, (f) $\mu_s\prime$ recovered with MC LUT and (g) $\mu_s\prime$ recovered with empirically derived LUT. Scale bar = 20mm.}
    \label{fig:BO_planar}
\end{figure}
Fig \ref{fig:BO_planar} shows the optical property maps generated for a segment of Barrett's oesophagus next to a segment of healthy oesophageal tissue. The tissue properties are designed to exhibit similar absorption coefficients, while the reduced scattering coefficient of the simulated BO is less than that of the adjacent healthy oesophageal tissue. At $635$ nm, the absorption coefficient of Barrett's oesophagus with mild chronic inflammation, $0.057$ mm$^{-1}$ is similar to that of healthy oesophageal tissue, while the reduced scattering coefficient of Barrett's oesophagus with mild chronic inflammation, $0.51$ mm$^{-1}$ is much less than that of healthy oesophageal tissue\cite{Sweer2019}. We simulated healthy oesophageal tissue as before with $A_{\rho} = 25$ and $S_{\rho} = 4236$, and Barrett's oesophagus with $A_{\rho} = 22$ and $S_{\rho} = 3215$. Fig \ref{fig:BO_planar}c,d,f and g show these optical properties are recovered as expected, demonstrating the capability of the simulation system to differentiate between tissue types.  We observe a close match between the Monte Carlo and empirically derived LUT results, which may arise due to the ideal geometric condition in which samples are totally flat, thereby reducing artefacts. We note that at the intersection region of the two simulated tissue types, there is a spike in both the optical properties, which results from effects at the interface and a small air gap that is present.  

\subsection{Simulation of optical property variation in tubular geometry}
\begin{figure}[!htpb]
    \centering
    \includegraphics[width=1\linewidth]{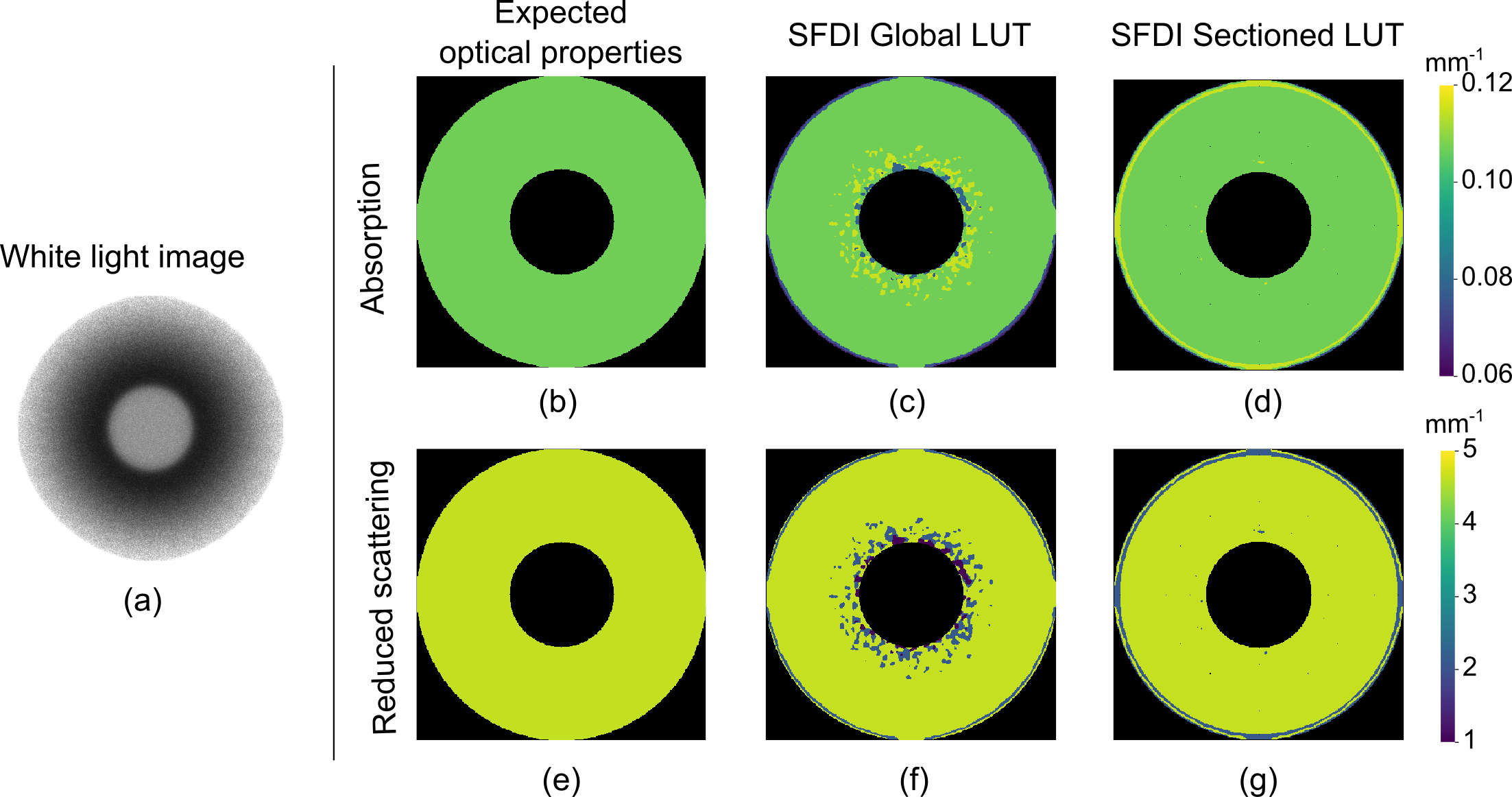}
    \caption{Comparison of sectioned an un-sectioned empirically derived LUT for tube wall material of $\mu_a = 0.107$ mm$^{-1}$ and $\mu_s \prime = 4.64$ mm$^{-1}$ (a) white light image of tube (b) expected absorption coefficient, $\mu_a$, (c) measured $\mu_a$ using un-sectioned LUT (d) measured $\mu_a$ using \emph{sectioned} LUT (e) expected reduced scattering coefficient, $\mu_s\prime$, (f) $\mu_s\prime$ measured using un-sectioned LUT, (g) $\mu_s\prime$ measured from \emph{sectioned} LUT. Tube inner diameter = 20mm.}
    \label{fig:Tube_lutvar}
\end{figure}

\begin{figure}
    \centering
    \includegraphics[width=0.9\linewidth]{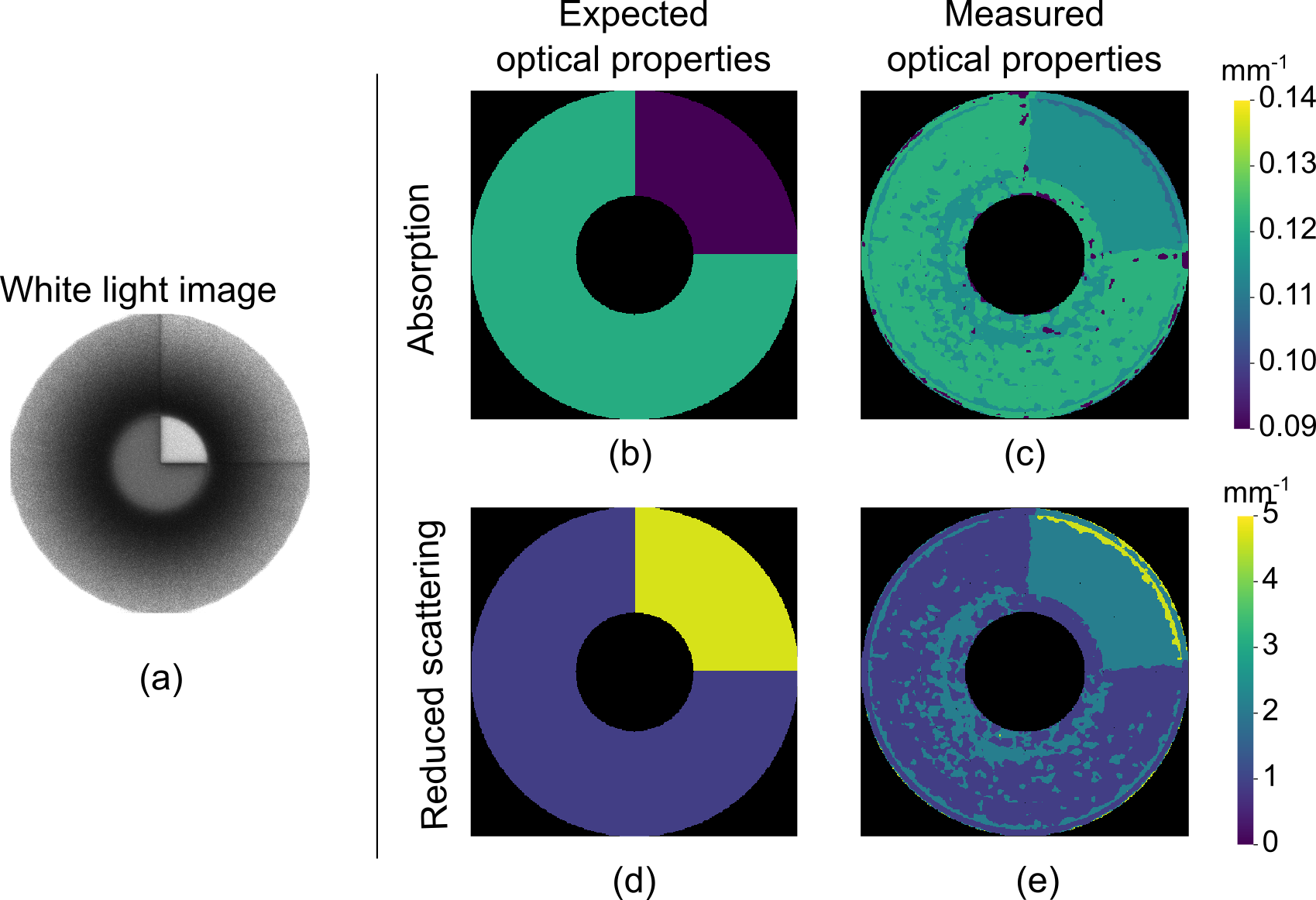}
    \caption{Imaging different material types within complex geometry, analogous to a lumen showing (a) white light image of the tube (b) expected absorption coefficient of tube material (c) measured absorption coefficient (d) expected reduced scattering coefficient of tube material and (e) measured reduced scattering coefficient. Tube inner diameter = 20mm.}
    \label{fig:Tube_var}
\end{figure}

The results presented here were taken using our modified projection pattern such that there was a uniform spatial frequency at the material surface enabling accurate optical property determination. This novel approach allows for the simulation of a geometry of any size to be generated and the desired illumination patterns of interest obtained for demodulation at the material surface. We initially recovered optical properties of the tube material via nearest neighbour interpolation using our Monte Carlo generated LUT described in Sect \ref{subsect:Calibration of material optical properties}. We use nearest-neighbour interpolation in this situation for robustness to points outside the convex hull of the LUT. This gives the results of Figure \ref{fig:Tube_lutvar} and \ref{fig:Tube_var} a quantized appearance. However, because it is relatively easy to generate more sample images using \emph{Blender}, the sample set could be expanded such that cubic interpolation could be reliably used with sufficient computing simulation time available. For the measurements here, the measured AC modulation amplitude was higher than expected, resulting in overestimation of scattering coefficients for low $S_{\rho}$ value materials. This difference arises because of the different geometry in the tubular case compared to the planar case. Here, the camera and projector are placed directly at the proximal end of the tube at no offset angle to one another, meaning the camera may be more prone to detecting specular reflections. This results in a higher AC modulation amplitude and reduced scattering coefficient offset.   

To overcome this issue, we introduced the empirically derived LUT described for tubular geometry in Sect \ref{subsect:Calibration of material optical properties}. Nearest neighbour interpolation was again used in this case. This was somewhat successful however it was not applicable along the length of the tube.  Significant improvement was achieved when sectioning the LUT as described in Sect \ref{subsec:ProjPatternTube}, shown in Fig \ref{fig:Tube_lutvar}. We calculated, over six varying material values, that the sectioned LUT method reduced the calculated absorption coefficient relative error by $50\%$ and reduced the calculated reduced scattering coefficient relative error from $8\%$ to $3\%$. 

Differentiating between optical properties inside lumen can be useful for detecting different diseases. To show the capability of the method to differentiate between optical properties inside a lumen, we simulated a tube with two different materials; three quarters had material properties $A_{\rho} = 100$ and $S_{\rho} = 5000$ for expected optical properties of $\mu_a = 0.121$ mm$^{-1}$ and $\mu_s \prime = 0.93$ mm$^{-1}$, and the top right quarter had material properties $A_{\rho} = 50$ and $S_{\rho} = 20000$ for expected optical properties $\mu_a = 0.066$ mm$^{-1}$ and $\mu_s \prime = 4.69$ mm$^{-1}$. The results are shown in Fig \ref{fig:Tube_var}. We note that the there is a distinct difference in material properties from the top right quarter and  the rest of the tube. However, there are artefacts present in the main part of the tube that do not match the optical properties. Due to the geometry of the system, it is likely that light is reflecting off multiple surfaces before reaching the camera which may contribute to these artefacts.

\section{Discussion}
These results demonstrate the capability of the \emph{Blender} system to simulate varying tissue types of various shapes, in different imaging geometries, and the capability of the SFDI imaging system to successfully determine their optical properties. This can be useful for many applications. Software such as \emph{OptogenSIM}\cite{OptogenSIM}, \emph{FullMonte}\cite{FullMonte} and  \emph{ValoMC}\cite{ValoMC} model Monte Carlo simulations in biologic
al relevant samples, however they suffer from a variety of limitations; such as incapability to generate realistic, complex sample geometries within the software and lack of full consideration for lighting conditions or camera positions. The presented SFDI simulation model can overcome many of the limitations of existing software by enabling custom configuration of illumination source and camera position and orientation, spatial frequency, and illumination pattern. This allows the introduction of real-world artefacts that help to test the limitations of a new system design.

SFDI can have various sources of errors\cite{Bodenschatz2014}, such as errors arising from assumptions made with selected light propagation model, differences in optical properties dependent on depth, divergence of the projection beam and how the spatial frequency may change with distance from projector to sample. These sources of error can be simulated in \emph{Blender} and their optimum solutions determined in a cost-effective, timely manner. Once solutions are found they can be applied to current systems to improve system accuracy.

We envisage the use of this system to simulate novel systems, as well as current systems, as a means to optimise an SFDI system to speed up the progression of system development and test the possibilities and limitations of the technique in different situations. 

Another potential application of this system could be to generate large data sets using SFDI, which may then be used in lieu of experimental data, eliminating the time and money necessary to go into data collection. These could be used to improve optical property uncertainty measurements by making large look up tables for specific system setups \cite{Pera2018}, to correct the optical property profile measurements using deep learning\cite{Aguenounon2020}, and has previously shown promise to train neural networks for SFDI demodulation\cite{Osman2022}.

\section{Conclusion}
We have further developed an SFDI and fringe profilometry imaging system in the open-source graphics software \emph{Blender} which enables the simulation of typical gastrointestinal conditions with specific absorption and reduced scattering coefficients in tubular imaging geometries relevant to that of the gastrointestinal tract. We have shown simulation of objects of specific shape, size and optical properties and successful imaging of these objects to recover maps of height, absorption and scattering. We anticipate our results will aid in the design of a future SFDI systems, e.g. miniaturised systems, by enabling the testing of different illumination geometries and patterns.

\begin{backmatter}
\bmsection{Funding}
The authors acknowledge support from a UKRI Future Leaders Fellowship (MR/T041951/1) and an ESPRC Ph.D. Studentship (2268555).

\bmsection{Acknowledgments}
We acknowledge open-source software resources offered by the Virtual Photonics Technology Initiative (https://virtualphotonics.org), at the Beckman Laser Institute, University of California, Irvine.

\bmsection{Disclosures}
The authors declare no conflicts of interest.

\bmsection{Data Availability Statement}
The data presented in this study are available from the the following source: [DOI to be inserted later].
%Discuss with George what data has to be made available
% We need to put all data in an online repository

\end{backmatter}

%%%%%%%%%%%%%%%%%%%%%%% References %%%%%%%%%%%%%%%%%%%%%%%%%

%%%%%%%%%% If using BibTeX:
%\bibliography{references}

\end{document}